
\magnification=\magstep 1
\baselineskip=20 pt
\font\bigbf = cmbx10 scaled\magstep 1
 1
 1

\line{}

\noindent

\centerline{\bigbf \hfill Influence of gauge-field fluctuations on \hfill}
\centerline{\bigbf \hfill composite fermions near the half-filled state
\hfill}

\vskip 1.0cm

\centerline{\hfill Yong Baek Kim, Patrick A. Lee, and
Xiao-Gang Wen \hfill}
\vskip 0.1cm
\centerline{\hfill \it Department of Physics, Massachusetts Institute of
Technology \hfill}
\centerline{\hfill \it Cambridge, Massachusetts 02139 \hfill}

\vskip 0.5cm

\centerline{\hfill P. C. E. Stamp \hfill}
\vskip 0.1cm
\centerline{\hfill \it Department of Physics, University of British Columbia
 \hfill}
\centerline{\hfill \it 6224 Agricultural Road, Vancouver, Canada V6T 1Z1
\hfill}

\vskip 0.5cm

\centerline{\hfill November 15, 1994 \hfill}

\vskip 0.5cm

\centerline{\hfill ABSTRACT \hfill}

\midinsert
\narrower
{\noindent\tenrm

Taking into account the transverse gauge field fluctuations, which
interact with composite fermions,
we examine the finite temperature compressibility of the fermions
as a function of an effective magnetic field
$\Delta B = B - 2 n_e hc/e$ ($n_e$ is the density of electrons)
near the half-filled state.
It is shown that, after including the lowest order gauge field
correction, the compressibility goes as
${\partial n \over \partial \mu} \propto e^{- \Delta \omega_c / 2 T}
\left ( 1 + {A (\eta) \over \eta - 1} {(\Delta \omega_c)^{2 \over 1 + \eta}
\over T} \right )$ for $T \ll \Delta \omega_c$, where
$\Delta \omega_c = {e \Delta B \over mc}$.
Here we assume that the interaction between the fermions is given by
$v ({\bf q}) = V_0 / q^{2 - \eta} \ (1 \le \eta \le 2)$,
where $A (\eta)$ is a $\eta$ dependent constant.
This result can be interpreted as a divergent correction to the
activation energy gap and is consistent with the divergent
renormalization of the effective mass of the composite fermions.

}
\endinsert

\vfill\vfill\vfill
\break

\centerline{\bf I. INTRODUCTION}

In 1989 Jiang {\it et al.} [1] observed that a two dimensional electron gas
in the fractional quantum Hall (FQH) regime,
at the filling fraction $\nu = 1/2$,
forms a metallic state even at very low temperatures.
At that time the only known
quantum metallic state at zero temperature was Landau's Fermi liquid
state in absence of a magnetic field.
Thus the experiment suggests that
the electrons at $\nu = 1/2$ may form a new quantum
metallic state at zero temperature.
The possibility of a new metallic state at $\nu = 1/2$ has
attracted a lot of attention [2-3].

On the theoretical side,
Jain [4] has introduced the composite fermion approach,
which successfully explains the stability of the sequence of the filling
fractions $\nu = p/(2p \pm 1)$ ($p$ is an integer).
Halperin, Lee, and Read (HLR) [5] observed that this sequence
is reminiscent of the Shubnikov-de Haas
oscillation of the conventional Fermi liquid in the presence of a weak
magnetic field, which indicates the possible existence of a Fermi
surface at $\nu = 1/2$.
This important observation and a set of new experiments [2,3] suggest
a strong connection between the Fermi liquid at zero magnetic field and
the new metallic state at $\nu =1/2$.
Using the Chern-Simons gauge field theory formulation of
the composite fermion approach, HLR realized these ideas and
developed a theory that describes the new metallic state [5].

A composite fermion is obtained by attaching an even number ($2n$) of flux
quanta to an electron and the transformation can be formally realized by
introducing an appropriate Chern-Simons gauge field [5-7].
At the mean field level, the FQH state with the filling factor
$\nu = p / (2np + 1)$ can be
described as the integral quantum Hall (IQH) state of the
composite fermions
with $p$ Landau levels occupied in an effective magnetic field
$\Delta B = B - B_{1/2n}$, where $B_{1/2n} = 2n n_e hc/e$ and $n_e$ is
the density of electrons.
Note that $\Delta B = 0$ for $\nu = 1/2n$ states so that,
within the mean field approximation, the composite fermions
can be described by the conventional Fermi liquid theory at these filling
factors [5,7].
In particular, the main sequence ($n=1$) of the hierarchical
structure of the FQH states [4-6] can be viewed as the
IQH effect of the composite fermions, which leads to the analogue of
the Shubnikov-de Haas effect near $\nu = 1/2$.

HLR had also gone beyond the mean field approximation
by including gauge field fluctuations within the random
phase approximation (RPA). This allowed them to construct
a modified Fermi-liquid-theory description of the $\nu = 1/2$ state.
They found that the gauge field fluctuations give rise to
a singular contribution to the self-energy in the one-particle
Green's function of the composite fermions [5,8].
Later various kinds of methods were used to go beyond
the perturbation theory [9-16], which were motivated by the fact that
the singular self-energy correction leads to a divergent effective mass
of the composite fermions [5].
If one applied this divergent effective mass to
the Shubnikov-de Haas effect near $\nu = 1/2$, one would
find that the gap $\Delta_p$ of $\nu = p/(2p \pm 1)$ FQH state goes down
faster than $1/p$ as $p \to \infty$:
$$ p \ \Delta_p \to 0 \hskip 1cm  \hbox{as} \hskip 1cm p \to \infty \ .
\eqno(1)$$
However, the one-particle Green's function of the composite
fermions is not gauge invariant.
Therefore, it is not clear whether the divergent effective mass in the
one-particle Green's function is related to the above energy gap
$\Delta_{p}$ which is measurable in real experiments.

There have been several studies of the two-particle correlation
functions which concentrated on the non-renormalization of the
gauge field propagator [11-14,16].
Recently three of us with Furusaki have
examined several gauge invariant two-particle
correlation functions for all ratios of $\omega$ and $q$ [17].
We found that, at low energies and in the long-wavelength limit,
the gauge field fluctuations do not cause any divergent correction
(up to two-loop level), and the two-particle correlation functions
have the Fermi-liquid forms with a finite effective mass if one
assumes a non-singular
Fermi-liquid-parameter-function $f_{\bf pp'}$ [17].
Fermi liquid form of the density-density correlation function in
the small $q$ and $\omega$ limit was also found in the eikonal
approximation [11] even though the result is not the same as
that of the two-loop perturbative calculation [17].
Altshuler, Ioffe, and Millis also examined the two-particle
correlation functions and especially found peculiar behaviors
near $q = 2 k_F$ [18].

We would like to mention that Fermi-liquid theory with a finite
effective mass is not the conclusive interpretation of the behaviors
of the density-density correlation function in the long wavelength
and the low frequency limit.
That is, it is still possible that
the effect of the divergent effective mass may be cancelled by a
contribution from a singular Fermi-liquid-parameter-function
$f_{\bf pp'}$ so that the density-density
correlation function for the long wavelength and the low energy limit
behaves as if the effective mass is finite [19].
Indeed, Stern and Halperin [19] calculated the energy gap
of the system from the one-particle Green's function of the
composite fermions in a finite effective magnetic field
$\Delta B$.
They argued that even though the one-particle Green's function is not
gauge-invariant, the edge of the spectral
function at zero temperature, across which the spectral function
vanishes, should be gauge-invariant.
By identifying the region where the spectral function vanishes,
they found an energy gap which is in agreement with the previous
self-consisteny treatment [5].
In view of the complexity of the problem, we feel that
it is important to investigate whether the effect of the large
enhancement of the effective mass will show up in some gauge-invariant
response functions.
In this paper, we calculate the lowest order correction (due to
the gauge field) to the finite temperature compressibility
as a function of an effective cyclotron frequency
$\Delta \omega_c = {e \Delta B \over m c}$
(where $m$ is the bare mass of
the fermions) in the limit of large $p$, {\it i.e.},
near $\nu = 1/2$.
We find that when a chemical potential $\mu$ lies exactly at the middle
of the successive effective Landau levels, for $T \ll \Delta \omega_c$,
the compressibility behaves as
$$
{\partial n \over \partial \mu} \propto e^{- \Delta \omega_c / 2 T}
\left ( 1 + {A (\eta) \over \eta - 1} {(\Delta \omega_c)^{2 \over 1 + \eta}
\over T} \right ) \ ,
\eqno{(2)}
$$
where $A (\eta)$ is a $\eta$-dependent positive dimensionful constant.
Here, we assume that the interaction between the fermions has the
form: $v({\bf q}) = V_0 / q^{2 - \eta} (1 \le \eta \le 2)$.
If we interpret the activation energy as a renormalized energy gap
$\Delta \omega^*_c$, {\it i.e.},
${\partial n \over \partial \mu} \propto e^{- \Delta \omega^*_c / 2 T}$,
it is given by
$\Delta \omega^*_c \approx \Delta \omega_c
\left ( 1 - {2 A (\eta) \over \eta - 1}
(\Delta \omega_c)^{-{\eta - 1 \over \eta + 1}} \right )$.
If we write $\Delta \omega^*_c = {e \Delta B \over m^* c}$,
the above result is consistent with a divergent correction to the
effective mass
$m^* / m \approx 1 + {2 A (\eta) \over \eta - 1}
(\Delta \omega_c)^{-{\eta - 1 \over \eta + 1}}$ because $A$ should
be proportional to a small expansion parameter, which is $1/N$ for
a large $N$ generalized model.
In particular, for the Coulomb interaction $(\eta = 1)$,
$m^* / m \approx 1 + 2 A (\eta=1) \
{\rm ln} \ (\epsilon_F / \Delta \omega_c)$
($\epsilon_F$ is the Fermi energy)
as predicted in terms of a self-consistent argument [5,19].

We would like to remark that a comparision with the recent
experimental measurements [3,20] of the
energy gap is complicated by the large impurity effects.
The disordered potential due to the impurities causes a
spatial fluctuation of the fermion density distribution, which
is equivalent to a large spatial fluctuation of the Chern-Simons
magnetic flux or $\Delta B$.
This means that there is a range of $\Delta B$ controlled by
the degree of disorder around the filling factor $\nu = 1/2$,
where impurity effects are very important.
In reality, this is the region where the gap measurement is not
possible due to the suppression of the amplitude of the
Shubnikov-de Haas effect.
We feel that a deeper understanding of the impurity effects is
necessary before
a recent experimental report [20] of an increase
in the effective mass near the boundary of the disorder
dominated region can be properly interpreted.

Before the main discussion, we would like to point out that
there is a gauge-invariant (for the Chern-Simons gauge field)
one-particle Green's function --- the Green's function of
the physical electrons,
which does not have a Fermi-liquid form [21] even though
the two-particle
Green's functions are similar to those of the Fermi liquid with a
finite or divergent effective mass.
In the first place, the electrons see a strong magnetic field and
the electron Green's function does not have any singularity at
$k_F$.
Secondly the spectral weight of the electron Green's function
is exponentially small at low energies even for the Coulomb
interaction, which is very different
from the Fermi liquid result [21].
Thus the $\nu = 1/2$ state really represents a
new class of metallic state.

The remainder of the paper is organized as follows.
In section II, we introduce the model and describe a method to
calculate the lowest order correction to the compressibility
${\partial n \over \partial \mu}$, where $n$ is the density
of the composite fermions.
in section III, the compressibility of the fermions is calculated
for $T \ll \Delta \omega_c \ll \mu$ when the chemical potential $\mu$
lies exactly at the middle of the two successive effective Landau levels.
In section IV, we discuss and contrast two different methods of
evaluating the compressibility and emphasize the gauge-invariant nature
of the method used in this paper.
We discuss and interpret our results in section V.

\vskip 0.5cm

\centerline{\bf II. THE MODEL AND THE COMPRESSIBILITY}

Let us consider the model for the composite fermions in which
a statistical gauge field or a Chern-Simons gauge field has been
introduced.
The model is given by [5,6]
$$
Z = \int \ D\psi \ D\psi^*  \ Da_{\mu} \ e^{\ i \int dt \ d^2 r \ {\cal L}} \ ,
\eqno{(3)}
$$
where the Lagrangian density ${\cal L}$ is
$$
\eqalign{
{\cal L} &= \psi^* (\partial_{0} + ia_0 - \mu) \psi - {1 \over 2m} \psi^*
(\partial_i - ia_i + i e A_i)^2 \psi \cr
&- {i \over 2 \pi {\tilde \phi}} \ a_0 \varepsilon^{ij} \partial_i a_j +
{1 \over 2} \int d^2 r' \ \psi^* ({\bf r}) \psi ({\bf r})
v ({\bf r} - {\bf r'}) \psi^* ({\bf r'}) \psi ({\bf r'}) \ ,
}
\eqno{(4)}
$$
where $\psi$ represents the fermion field and ${\tilde \phi}$ is an even
number $2n$ which is the number of
flux quanta attached to an electron, and $v ({\bf r}) \propto V_0 / r^{\eta}$
is the Fourier transform of
$v ({\bf q}) = V_0 / q^{2 - \eta} \ (1 \le \eta \le 2)$ which denotes
the interaction between the fermions.
We choose the Coulomb gauge $\nabla \cdot {\bf a} = 0$.
Note that the integration over $a_0$ enforces the following constraint [5,6]:
$$
\nabla \times {\bf a} = 2 \pi {\tilde \phi} \
\psi^* ({\bf r}) \psi ({\bf r}) \ .
\eqno{(5)}
$$
The saddle point of the above action is given by the following
conditions [5,6]:
$$
\nabla \times {\bf a} = 2 \pi {\tilde \phi} \ n_e \equiv e B_{1/2n}
\ \ {\rm and} \ \ a_0 = 0 \ .
\eqno{(6)}
$$

Therefore, at the mean field level, the fermions see an effective magnetic
field ($\Delta {\bf A} \equiv {\bf A} - {\bf a}/e$):
$$
\Delta B = \nabla \times \Delta {\bf A} = B - B_{1/2n} \ ,
\eqno{(7)}
$$
which becomes zero at the Landau level filling factor $\nu = 1/2n$.
The IQH effect of the fermions may appear when the effective
Landau level filling factor $p = {n_e h c \over e \Delta B}$ becomes
an integer.
This implies that the real external magnetic field is given by
$B = B_{1/2n} + \Delta B = {n_e h c \over e} \ \left (
{2np + 1 \over p} \right )$ which corresponds to a FQH state
of electrons with the filling factor $\nu = {p \over 2np +1}$ [5,6].

After integrating out the fermions and including gauge field fluctuations
within the random phase approximation, the effective action of
the gauge field can be obtained [5]
$$
S_{\rm eff} = {1 \over 2} \int {d^2 q \over (2 \pi)^2} \
{d\omega \over 2 \pi}
\ \delta a^*_{\mu} ({\bf q}, \omega) \
D^{-1}_{\mu \nu} ({\bf q},\omega,\Delta \omega_c)
\ \delta a_{\nu} ({\bf q}, \omega) \ ,
\eqno{(8)}
$$
where $D^{-1}_{\mu \nu} ({\bf q},\omega,\Delta \omega_c)$ was calculated
by several authors [5,6,22-24].
For our purpose, the $2 \times 2$ matrix form for $D^{-1}_{\mu \nu}$
is sufficient so that
$\mu, \nu = 0, 1$ and $1$ represents the direction that is
perpendicular to ${\bf q}$ [5].

The compressibility of the fermions ${\partial n \over \partial \mu}
(\mu, \Delta \omega_c)$ as a function of chemical potential $\mu$
and an effective cyclotron frequency
$\Delta \omega_c = {e \Delta B \over mc}$ can be obtained from
$n (\mu, \Delta \omega_c) = - {\partial \Omega \over \partial \mu}$
($n$ is the density of the fermions),
{\it i.e.}, ${\partial n \over \partial \mu} = - {\partial^2 \Omega \over
\partial \mu^2}$.
The density of the free fermions $n_0 (\mu, \Delta \omega_c)$ and
the lowest order correction $n_1 (\mu, \Delta \omega_c)$ due to the
transverse part of the gauge field fluctuations are given by the
diagrams in Fig.1 (a) and (b) respectively.
These contributions can be obtained from the relations
$n_0 (\mu, \Delta \omega_c) = - {\partial \Omega_0 \over \partial \mu}$
and $n_1 (\mu, \Delta \omega_c) = - {\partial \Omega_1 \over \partial \mu}$,
where $\Omega_0$ and $\Omega_1$ are the thermodynamic potential of the free
fermions and the lowest order correction to the thermodynamic potential given
by the diagrams in Fig.3 (a) and (b) respectively.

The density of the free fermions $n_0 (\mu, \Delta \omega_c)$ at finite
temperatures can be written as
$$
n_0 (\mu, \Delta \omega_c) = {m (\Delta \omega_c) \over 2\pi} \sum_{l}
n_F (\xi_l) \ ,
\eqno{(9)}
$$
where $\xi_{l} = (l + 1/2) (\Delta \omega_c) - \mu$ and
$n_F (x) = {1 \over e^{x/T} + 1}$.
Thus the compressibility of the free fermions is given by
$$
{\partial n_0 \over \partial \mu} =
{m \over 2 \pi} \ {\Delta \omega_c \over T} \sum_{l}
n_F (\xi_l) (1 - n_F (\xi_l)) \ .
\eqno{(10)}
$$

The lowest order correction (due to the transverse part of the
gauge field) to the density of the fermions can be obtained from
$$
n_1 (\mu, \Delta \omega_c) = T \sum_{i \nu_n} \sum_{\bf q}
D_{11} ({\bf q}, i \nu_n) \ {\partial \over \partial \mu}
\Pi_{11} ({\bf q}, i \nu_n) \ ,
\eqno{(11)}
$$
where $\nu_n = 2 \pi n T$ is the Matsubara frequency.
Here $\Pi_{11}$ is the transverse part of the fermion polarization
bubble:
$$
\Pi_{11} ({\bf q}, i \nu_n) =
- \sum_{lm} |M_{lm} ({\bf q})|^2 \ {n_F (\xi_l) - n_F (\xi_m) \over
i \nu_n - \xi_m + \xi_l} -
{1 \over m} \left ( {m \Delta \omega_c \over 2 \pi}
\sum_{l} n_F (\xi_l) \right ) \ ,
\eqno{(12)}
$$
where $|M_{lm} ({\bf q})|^2$ comes from the form of the
current-current vertex and is calculated by several authors [6,22,23].
After analytic continuation $i \nu_n \rightarrow \nu + i 0^{+}$,
one gets the real part and the imaginary part of the retarded polarization
function:
$$
\eqalign{
\Pi^{'}_{11} ({\bf q}, \nu) &=
- \sum_{lm} |M_{lm} ({\bf q})|^2 \ {n_F (\xi_l) - n_F (\xi_m) \over
\nu - \xi_m + \xi_l} -
{1 \over m} \left ( {m \Delta \omega_c \over 2 \pi}
\sum_{l} n_F (\xi_l) \right ) \ , \cr
\Pi^{''}_{11} ({\bf q}, \nu) &=
\pi \sum_{lm} |M_{lm} ({\bf q})|^2 \ \left [ \ n_F (\xi_l) - n_F (\xi_m)
\ \right ] \ \delta (\nu - \xi_m + \xi_l) \ .}
\eqno{(13)}
$$
Here we use the convention that $A^{'}$ and $A^{''}$ represent
the real and the imaginary parts of a quantity $A$.
Now the correction to the compressibility can be obtained as
$$
{\partial n_1 \over \partial \mu} =
T \sum_{i \nu_n} \sum_{\bf q} \left [ \
D_{11} ({\bf q}, i \nu_n) \ {\partial^2 \over \partial \mu^2}
\Pi_{11} ({\bf q}, i \nu_n) \ + \
{\partial \over \partial \mu} D_{11} ({\bf q}, i \nu_n)
\ {\partial \over \partial \mu}
\Pi_{11} ({\bf q}, i \nu_n) \ \right ] \ .
\eqno{(14)}
$$

For calculational convenience, we introduce
${\widetilde D}_{11} ({\bf q}, i \nu_n)$ which does not depend on
$\mu$.
Then the correction to the physical fermion density
$n_1 (\mu, \Delta \omega_c)$ can be obtained from
$n_1 (\mu, \Delta \omega_c) = - {\partial \Omega_{\rm toy} \over
\partial \mu}$, where
$$
\Omega_{\rm toy} = T \sum_{i \nu_n} \sum_{\bf q}
{\widetilde D}_{11} ({\bf q}, i \nu_n) \ \Pi_{11} ({\bf q}, i \nu_n) \ ,
\eqno{(15)}
$$
and replace ${\widetilde D}_{11} ({\bf q}, i \nu_n)$ by
$D_{11} ({\bf q}, i \nu_n)$ after taking the derivative with respect
to $\mu$.
Using the spectral representation, one can write $\Omega_{\rm toy}$ as
$$
\eqalign{
\Omega_{\rm toy} &= \Omega_a + \Omega_b \ , \cr
\Omega_a &= \sum_{\bf q} \int^{\infty}_{-\infty} {dx \over \pi} \
n_B (x) \ {\widetilde D}^{'}_{11} ({\bf q}, x) \
\Pi^{''}_{11} ({\bf q}, x) \ , \cr
\Omega_b &= \sum_{\bf q} \int^{\infty}_{-\infty} {dx \over \pi} \
n_B (x) \ {\widetilde D}^{''}_{11} ({\bf q}, x) \
\Pi^{'}_{11} ({\bf q}, x) \ ,}
\eqno{(16)}
$$
where $n_B (x) = {1 \over e^{x/T} - 1}$.
After taking the derivative with respect to $\mu$ and replacing
${\widetilde D}_{11}$ by $D_{11}$, we get the lowest order correction to
the density of the fermions:
$$
\eqalign{
n_1 &= n_a  + n_b  \ , \cr
n_a &= -
\sum_{\bf q} \int^{\infty}_{-\infty} {dx \over \pi} \
n_B (x) \ D^{'}_{11} ({\bf q}, x) \
{\partial \over \partial \mu} \Pi^{''}_{11} ({\bf q}, x) \ , \cr
n_b &= -
\sum_{\bf q} \int^{\infty}_{-\infty} {dx \over \pi} \
n_B (x) \ D^{''}_{11} ({\bf q}, x) \
{\partial \over \partial \mu} \Pi^{'}_{11} ({\bf q}, x) \ .}
\eqno{(17)}
$$

For the lowest order correction ${\partial n_1 \over \partial \mu}$
to the compressibility, the derivative with respect to $\mu$
should be taken for both $D_{11}$ and $\Pi_{11}$.
Thus ${\partial n_1 \over \partial \mu}$ can be written as
$$
\eqalign{
{\partial n_1 \over \partial \mu} &=
{\partial n_a \over \partial \mu} +
{\partial n_b \over \partial \mu} \ , \cr
{\partial n_a \over \partial \mu} &= -
\sum_{\bf q} \int^{\infty}_{-\infty} {dx \over \pi} \
n_B (x) \ \left [ \ D^{'}_{11} {\partial^2 \Pi^{''}_{11} \over \partial \mu^2}
+ {\partial D^{'}_{11} \over \partial \mu}
{\partial \Pi^{''}_{11} \over \partial \mu} \ \right ] \ , \cr
{\partial n_b \over \partial \mu} &= -
\sum_{\bf q} \int^{\infty}_{-\infty} {dx \over \pi} \
n_B (x) \ \left [ \ D^{''}_{11} {\partial^2 \Pi^{'}_{11} \over \partial \mu^2}
+ {\partial D^{''}_{11} \over \partial \mu}
{\partial \Pi^{'}_{11} \over \partial \mu} \ \right ] \ .}
\eqno{(18)}
$$
Note that Eq.(18) is equivalent to Eq.(14).
This procedure generates the diagrams for the compressibility,
which are shown in Fig.4.
In the next section, we evaluate the expressions for the
compressibility.

\vskip 0.5cm

\centerline{\bf III. THE FINITE TEMPERATURE COMPRESSIBILITY FOR
$T \ll \Delta \omega_c \ll \mu$}

In this section, we calculate the compressibility of the fermions
as a function of $\Delta \omega_c$ and $T$ in the limit
$T \ll \Delta \omega_c$.
First we would like to give a general discussion of
the interaction effects on the compressibility.
For free fermions at zero temperature and finite magnetic field,
${d n \over d \mu} = \sum_{m} \delta (\mu - (n+{1 \over 2}))
\Delta \omega_c$ is the density of states.
Each $\delta$-function corresponds to a
degenerate effective Landau level.
The interaction has two kinds of effects
on the compressibility ${d n \over d \mu}$.
First, the interaction effects split the degeneracy of the states
in each effective Landau level (when the effective Landau level
is partially filled).
This effect spreads the $\delta$-function in the free fermion
compressibility into broadened peaks.
The width of the peak (defined as the width of the region
where ${d n \over d \mu} \not= 0$) can be viewed as the width
of the effective Landau bands ({\it i.e.}, the
broadened effective Landau levels).
Second, the interaction effects
may shift the center of the effective Landau bands.
However, since the average
compressibility over many effective Landau levels
is not changed by the transverse
gauge field interaction,
we expect that such an interaction can only cause
a uniform shift of the center of the effective Landau bands,
as one can see later in our explicit calculations.
The activation energy gap measured in the transport experiments
is given by the gap between the effective Landau bands.
Thus the uniform shift is not important for the calculation of
of the experimentally measurable activation energy gap.
In the following calculations,
we will assume that the chemical potential $\mu$ lies exactly at
the middle of the two successive effective Landau levels,
and investigate the activated behavior of the compressibility.
In this case, the uniform shift of the center of the effective
Landau bands is cancelled out and
does not appear in the compressibility.

Let $p$ be the number of filled effective Landau levels.
For the free fermions, when $T \ll \Delta \omega_c$, we can expect
that the compressibility shows a thermally activated behavior.
In fact, from Eq.(10) and for $T \ll \Delta \omega_c$,
it can be shown that at finite temperatures
the compressibility of the free fermions can be written as
$$
{\partial n_0 \over \partial \mu} =
{m \over 2\pi} \ {\Delta \omega_c \over T}
\left ( e^{-|\xi_p|/T} + e^{-\xi_{p+1}/T} \right )
\ + \ {\cal O} (e^{-2|\xi_p|/T}) \ .
\eqno{(19)}
$$
Note that it becomes
$$
{\partial n_0 \over \partial \mu} =
{m \Delta \omega_c \over \pi T} \ e^{-\Delta \omega_c / 2T}
\ + \ {\cal O} (e^{-\Delta \omega_c / T}) \
\eqno{(20)}
$$
for a chemical potential lying exactly
at the middle of the Landau levels labeled by $p$ and $p+1$.
Our aim is to calculate the lowest order correction (due to the gauge field
fluctuations) to the above free fermion result.

In order to calculate the lowest order correction
${\partial n_1 \over \partial \mu}$,
we consider first $\Omega_{\rm toy} = \Omega_a + \Omega_b$.
Substituting Eq.(13) to Eq.(16), we get
$$
\eqalign{
\Omega_a &= \Omega_{a1} + \Omega_{a2} \ , \cr
\Omega_{a1} &= \sum_{\bf q} \sum_{l} |M_{ll} ({\bf q})|^2 \
{\widetilde D}^{'}_{11} ({\bf q}, 0) \ n_F (\xi_l) (1 - n_F (\xi_l)) \ , \cr
\Omega_{a2} &= \sum_{\bf q} \sum_{l \not= m} |M_{lm} ({\bf q})|^2 \
{\widetilde D}^{'}_{11} ({\bf q}, \xi_m - \xi_l) \
n_F (\xi_m) (1 - n_F (\xi_l)) \ ,}
\eqno{(21)}
$$
and
$$
\eqalign{
\Omega_b &= \Omega_{b1} + \Omega_{b2} \ , \cr
\Omega_{b1} &= \sum_{\bf q} \int^{\infty}_0 {dx \over \pi} \
(1 + 2 n_B (x)) \ {\widetilde D}^{''}_{11} ({\bf q}, x)
\left [ - \sum_{lm} |M_{lm} ({\bf q})|^2 \ {n_F (\xi_l) - n_F (\xi_m) \over
x - \xi_m + \xi_l} \right ] \ , \cr
\Omega_{b2} &= \sum_{\bf q} \int^{\infty}_0 {dx \over \pi} \
(1 + 2 n_B (x)) \ {\widetilde D}^{''}_{11} ({\bf q}, x)
\left [ - {\Delta \omega_c \over 2 \pi}
\sum_{l} n_F (\xi_l) \right ] \ .}
\eqno{(22)}
$$
Now some explanations for each contribution are in order.
$\Omega_{a1}$ and $\Omega_{a2}$ are contributions from
the exchange interaction via the gauge field and
represent the effect of the intra-Landau level and the
inter-Landau level particle-hole excitations respectively.
$\Omega_{b1}$ and $\Omega_{b2}$ are due to the thermal
and the quantum (represented by $n_B (x)$ and $1$ in the
factor $1 + 2 n_B (x)$) fluctuations of the gauge field.
Note that the quantum contribution survives in the
$T \rightarrow 0$ limit.
In particular, $\Omega_{b2}$ comes from the diamagnetic
coupling between the fermions and the gauge field.
We also note that the intra-Landau level terms (with $l=m$)
are associated with the splitting of the degenerate states
in each Landau-level, and contribute to the spreading
of the Landau-levels.
On the other hand, the inter-Landau level terms (with $l \not= m$) will
contribute to the shift of the center of the Landau bands.

The corresponding contributions to the density of the fermions
are defined as $n_{a} = - {\partial \Omega_{a} \over \partial \mu}$
and $n_{b} = - {\partial \Omega_{b} \over \partial \mu}$.
Thus the correction to the density of the fermions ${\partial n_1
\over \partial \mu}$ is given by
${\partial n_1 \over \partial \mu} = {\partial n_{a} \over \partial \mu}
+ {\partial n_{b} \over \partial \mu}$.
Now we are going to find the contributions which are order of
$e^{-|\xi_p|/T}$ or $e^{-\xi_{p+1}/T}$.
Note that
${\partial n_{a} \over \partial \mu} = {\partial n_{a1} \over \partial \mu}
+ {\partial n_{a2} \over \partial \mu}$, where
$n_{a1} = - {\partial \Omega_{a1} \over \partial \mu}$ and
$n_{a2} = - {\partial \Omega_{a2} \over \partial \mu}$.
In the appendix, we show that ${\partial n_{a2} \over \partial \mu}$
is order of $e^{-2|\xi_p|/T}$ which is exponentially smaller
than $e^{-|\xi_p|/T}$ or $e^{-\xi_{p+1}/T}$.
It is also shown that ${\partial n_{b} \over \partial \mu}$
is order of $e^{-2|\xi_p|/T}$ after a partial cancellation between
$\Omega_{b1}$ and $\Omega_{b2}$ by the f-sum rule.

Now let us look at ${\partial n_{a1} \over \partial \mu}$ for which
a detailed expression is given in the appendix.
As mentioned before, we assume that we are very close to the
half-filled state, {\it i.e.}, $\mu / \Delta \omega_c \gg 1$, which
also corresponds to the large $p$ limit.
In this case, it can be shown that
$$
\eqalign{
{\partial n_{a1} \over \partial \mu} &\approx
- {1 \over T^2} \ \sum_{\bf q} \ \left [  e^{-\xi_{p+1}/T} \
|M_{p+1 p+1} ({\bf q})|^2 \ D^{'}_{11} ({\bf q}, 0) \ + \
e^{-|\xi_{p}|/T} \
|M_{p p} ({\bf q})|^2 \ D^{'}_{11} ({\bf q}, 0) \right ] \cr
&\hskip 0.7cm + \ {\cal O} (e^{-2|\xi_{p}|/T}) \cr
&\approx
- {1 \over T^2} \left [  e^{-\xi_{p+1}/T} + e^{-|\xi_{p}|/T} \right ] \
\sum_{\bf q} |M_{p p} ({\bf q})|^2 \ D^{'}_{11} ({\bf q}, 0) \ + \
{\cal O} (e^{-2|\xi_{p}|/T}) \ .}
\eqno{(23)}
$$
For $\xi_{p+1} = |\xi_p| = \Delta \omega_c / 2$, we get
$$
{\partial n_{a1} \over \partial \mu} \approx
- {2 \over T^2} \ e^{-\Delta \omega_c / 2T} \
\sum_{\bf q} |M_{p p} ({\bf q})|^2 \ D^{'}_{11} ({\bf q}, 0) \ + \
{\cal O} (e^{-\Delta \omega_c / T}) \ .
\eqno{(24)}
$$
Thus ${\partial n_1 \over \partial \mu} =
{\partial n_{a1} \over \partial \mu} \ + \
{\cal O} (e^{-\Delta \omega_c / T})$.

Now let us evaluate the following quantity.
$$
I = \sum_{\bf q} |M_{p p} ({\bf q})|^2 \ D^{'}_{11} ({\bf q}, 0)
= 2 \sum_{\bf q} |M_{p p} ({\bf q})|^2 \int^{\infty}_0 {dy \over \pi} \
{D^{''}_{11} ({\bf q}, y) \over y} \ .
\eqno{(25)}
$$
Note that the matrix element $|M_{p p} ({\bf q})|^2$ comes from
the vertex of the paramagnetic part of the current-current correlation
function.
For the large $p$ limit or $\mu / \Delta \omega_c \gg 1$,
we may use a semiclassical approximation
$j \approx v_F \rho$, where $j$ and $\rho$ are the current and the density
of the fermions.
Thus $|M_{p p} ({\bf q})|^2$ can be approximated as
$|M_{p p} ({\bf q})|^2 \approx v^2_F |M^{00}_{p p} ({\bf q})|^2$,
where $|M^{00}_{p p} ({\bf q})|^2$ is the corresponding matrix element
for the density-density correlation function [6,22-24].
Using the above approximation, we get
$$
|M_{p p} ({\bf q})|^2 \approx
{v^2_F \over 2 \pi l^2_c} \ e^{-X} \ \left [ L^{0}_p (X) \right ]^2 \ ,
\eqno{(26)}
$$
where $l^2_c \equiv {\hbar c \over e \Delta B}$,
$X \equiv {1 \over 2} q^2 l^2_c$, and $L^{0}_p (X)$ is a
Laguerre polynomial.
For the large $p$ limit, $L^{\alpha}_p (X)$ can be approximated as [25]
$$
L^{\alpha}_p (X) \approx
{1 \over \pi} \ e^{X / 2} \
X^{- {\alpha \over 2} - {1 \over 4}} \
p^{{\alpha \over 2} - {1 \over 4}} \
{\rm cos} \left ( 2 \sqrt{p X} - {\alpha \pi \over 2} - {\pi \over 4}
\right ) \ .
\eqno{(27)}
$$
We use $p \approx \mu / \Delta \omega_c$ and the above results to get
$$
|M_{p p} ({\bf q})|^2 \approx {m v_F \over \pi^3} \
{(\Delta \omega_c)^2 \over q} \
{\rm cos}^2 \left ( \sqrt{2 p} \ q l_c - {\pi \over 4} \right ) \ .
\eqno{(28)}
$$

Note that $D^{''}_{11} ({\bf q}, y)$ consists of two contributions
coming from the intra-Landau-level and the inter-Landau-level
processes respectively.
That is, in the particle-hole bubbles appearing in the $1/N$ expansion
(or the RPA approximation) of the gauge field propagator, the particle line
and the hole line may carry the same effective-Landau-level index or
different indices.
For the inter-Landau-level process, there is an excitation gap which
is the order of $\Delta \omega_c$.
Thus, for $y < \Delta \omega_c$, the intra-Landau-level
process is the only contribution to $D^{''}_{11} ({\bf q}, y)$.
As shown before, the intra-Landau-level contribution to a particle-hole
bubble gives rise to the $n_F (\xi_l) (1 - n_F (\xi_l))$ factor in
the gauge field propagator, which becomes
exponentially small for $T \ll \Delta \omega_c$.
This suggests that
$D^{''}_{11} ({\bf q}, y)$ becomes exponentially small
for $y < \Delta \omega_c$ and $T \ll \Delta \omega_c$ so that
we can ignore the contribution coming from $y < \Delta \omega_c$
for our purpose.
Thus we consider only the contribution coming from the
inter-Landau-level process, which appears only above the
gap $\Delta \omega_c$.
For $\mu / \Delta \omega_c \gg 1$ or the large $p$ approximation,
one may argue that the smearing of the discrete spectral function
$D^{''}_{11} ({\bf q}, y)$ of the gauge field propagator, which comes
from the Landau-level structure, does not cause any significant
change in the global behavior of the response functions.
Therefore, we use $D^{''}_{11} ({\bf q}, y)$ for
$\Delta B = 0$ instead of $D^{''}_{11} ({\bf q}, y)$
for finite $\Delta B$, but a lower cutoff $\Delta \omega_c$ is
introduced in the $y$ integral in Eq.(25)
to mimic the gap in $D^{''}_{11} ({\bf q}, y)$.
Since the precise value of the gap is not known, the numerical
coefficient of the final answer to the response function
is unreliable, but
the functional dependence on $\Delta \omega_c$ is not affected.

The transverse gauge field propagator $D_{11} ({\bf q}, \omega)$
for $\Delta B = 0$ is given by
$1 / (- i \gamma {\omega \over q} + \chi q^{\eta})$ [5], where
$\gamma = {2 n_e \over k_F}$, $\chi = {1 \over 24 \pi m} +
{V_0 \over (2 \pi {\tilde \phi})^2}$ for $\eta = 2$, and
$\chi = {V_0 \over (2 \pi {\tilde \phi})^2}$ for $\eta \not= 2$.
For the large p limit, evaluation of the $q$ integral in Eq.(25)
gives us
$$
\int {d^2 q \over (2 \pi)^2} \ |M_{p p} ({\bf q})|^2 \
D^{''}_{11} ({\bf q}, y)
\approx
- {m v_F \over 8 \pi^3} \ {1 \over 1 + \eta} \ {1 \over {\rm sin} \
\left ( {\pi \over 1 + \eta} \right )} \ \gamma^{-{\eta-1 \over \eta + 1}}
\ \chi^{-{2 \over 1 + \eta}} \ y^{-{\eta-1 \over \eta + 1}} \
(\Delta \omega_c)^2 \ .
\eqno{(29)}
$$
Now we can perform the $y$ integral, yielding
$$
\eqalign{
I &\approx 2 \int^{\infty}_{\Delta \omega_c} {dy \over \pi} \
\sum_{\bf q} |M_{p p} ({\bf q})|^2 \ {D^{''}_{11} ({\bf q}, y) \over y} \cr
&=
- {m v_F \over 4 \pi^4} \ {1 \over \eta-1} \ {1 \over {\rm sin} \
\left ( {\pi \over 1 + \eta} \right )} \ \gamma^{-{\eta-1 \over \eta + 1}}
\ \chi^{-{2 \over 1 + \eta}} \ (\Delta \omega)^{\eta + 3 \over \eta +1} \ .}
\eqno{(30)}
$$
Therefore, for $\xi_{p+1} = |\xi_p| = \Delta \omega_c / 2$, we get
$$
{\partial n_1 \over \partial \mu} \approx
{A (\eta) \over \eta -1} \ {m \over \pi} \
{(\Delta \omega_c)^{\eta + 3 \over \eta + 1} \over T^2} \ ,
\eqno{(31)}
$$
where
$$
A (\eta) = {v_F \over 2 \pi^3} \ {1 \over {\rm sin} \
\left ( {\pi \over 1 + \eta} \right )} \ \gamma^{-{\eta-1 \over \eta + 1}}
\ \chi^{-{2 \over 1 + \eta}} \ .
\eqno{(32)}
$$
Combining the result of Eq.(31) and that of the free fermions given by
Eq.(20), we get
$$
{\partial n \over \partial \mu} \approx {m (\Delta \omega_c) \over \pi T} \
e^{-\Delta \omega_c / 2T} \left ( 1 + {A (\eta) \over \eta - 1} \
{(\Delta \omega_c)^{2 \over 1 + \eta} \over T} \right ) \ .
\eqno{(33)}
$$
This is the central result of this paper.

Note that $A (\eta)$ should be proportional to a small expansion
parameter, for example, $1/N$ in a large $N$ generalized model.
Thus $1 + {A (\eta) \over \eta - 1} \
{(\Delta \omega_c)^{2 \over 1 + \eta} \over T} \approx
e^{{A (\eta) \over \eta - 1} \
{(\Delta \omega_c)^{2 \over 1 + \eta} \over T}}$ so that
the result of Eq.(33) is consistent with the renormalized energy
gap $\Delta \omega^*_c \approx \Delta \omega_c
\left ( 1 - {2 A (\eta) \over \eta - 1}
(\Delta \omega_c)^{-{\eta - 1 \over \eta + 1}} \right )$ if we write
$\partial n / \partial \mu \propto e^{-\Delta \omega^*_c / 2T}$.
This implies that $m^* / m \approx 1 + {2 A (\eta) \over \eta - 1}
(\Delta \omega_c)^{-{\eta - 1 \over \eta + 1}}$ from
$\Delta \omega^*_c = {e \Delta B \over m^* c}$.
In particular, for the Coulomb interaction $(\eta = 1)$,
$\Delta \omega^*_c \approx \Delta \omega_c
\left ( 1 - 2 A (\eta=1) \ {\rm ln} \
(\epsilon_F / \Delta \omega_c) \right )$ and
$m^* / m \approx 1 + 2 A (\eta=1) \
{\rm ln} \ (\epsilon_F / \Delta \omega_c)$.
These results were predicted by HLR in terms of a
self-consistency argument [5] and are also consistent with
the recent work of Stern and Halperin [19].

\vskip 0.5cm

\centerline{\bf IV. POLARIZATION BUBBLE VERSUS SELF-ENERGY}

In the previous sections, we used Eq.(16) and the subsequent
derivatives of $\Omega_{\rm toy}$ to get the correction to
the compressibility.
There is an alternative way to express $\Omega_{\rm toy}$,
which involves the use of the self-energy.
That is, Eq.(15) can be written as
$$
\Omega_{\rm toy} = - T \sum_{i \omega_n} \sum_{l}
{m \Delta \omega_c \over 2 \pi} \
\Sigma (\xi_{l}, i \omega_n) \ G (\xi_{l}, i \omega_n) \ ,
\eqno{(34)}
$$
where $\omega_n = (2n + 1) \pi T$ is the Matsubara frequency and
$G (\xi_{l}, i \omega_n) = {1 \over i \omega_n - \xi_{l}}$.
$\Sigma (\xi_{l}, i \omega_n)$ is the one-loop self-energy
correction dressed by the gauge field ${\widetilde D}_{11}$ and
is given by the diagrams in Fig.5.
We note that $\Omega_{\rm toy}$ is finite, whereas $\Sigma$ is
known to be infinite (for $\eta = 2$) at finite temperatures
and $\Delta \omega_c = 0$.
In this section, we wish to clarify how this apparent difficulty
is resolved.
Using the spectral representation, we can rewrite Eq.(34) as
$$
\eqalign{
\Omega_{\rm toy} &= \Omega_c + \Omega_d \ , \cr
\Omega_c &= {m \Delta \omega_c \over 2 \pi}
\sum_{l} \int^{\infty}_{-\infty} {dx \over \pi} \
n_F (x) \ \Sigma^{''} (\xi_{l}, x) \
G^{'} (\xi_{l}, x) \ , \cr
\Omega_d &= {m \Delta \omega_c \over 2 \pi} \sum_{l}
\int^{\infty}_{-\infty} {dx \over \pi} \
n_F (x) \ \Sigma^{'} (\xi_{l}, x) \ G^{''} (\xi_{l}, x) \ ,}
\eqno{(35)}
$$

Now we would like to compare two ways of calculating
$\Omega_{\rm toy}$.
First let us discuss the case of $\Delta \omega_c = 0$.
If we use Eq.(16), one can show that $\Omega_{a}$ is finite
by using $\Pi^{''}_{11} ({\bf q}, x) \approx - \gamma x / q$.
Suppose that we are going to use only the first diagram of the
transeverse part of the polarization bubble in Fig.2 (b)
to calculate $\Omega_{b}$.
Since the leading contribution of the first diagram to $\Pi^{'}$
is given by $n_0 / m$ where $n_0$ is the density of the free fermions,
it can be shown that $\Omega_{b}$ diverges in this case.
However, the second diagram also contributes $- n_0 / m$ which
cancels the constant term of the first diagram.
This cancellation is required by the gauge-invariance.
As a result, $\Pi^{'} \approx \chi_0 q^2$ with
$\chi_0 = {1 \over 24 \pi m}$ so that $\Omega_{b}$ becomes finite.
In particular, for the short range interaction ($\eta = 2$),
$\Omega_{a}$ and $\Omega_{b}$ give rise to the same contributions
with different coefficients.

Next we examine what happens if we use Eq.(35) which expresses
$\Omega_{\rm toy}$ in terms of the self-energy.
For $\Delta \omega_c = 0$, Eq.(35) can be rewritten as
$$
\eqalign{
\Omega_{\rm toy} &= \Omega_c + \Omega_d \ , \cr
\Omega_c &= \sum_{\bf k} \int^{\infty}_{-\infty} {dx \over \pi} \
n_F (x) \ \Sigma^{''} (\xi_{\bf k}, x) \
G^{'} (\xi_{\bf k}, x) \ , \cr
\Omega_d &= \sum_{\bf k}
\int^{\infty}_{-\infty} {dx \over \pi} \
n_F (x) \ \Sigma^{'} (\xi_{\bf k}, x) \ G^{''} (\xi_{\bf k}, x) \ .}
\eqno{(36)}
$$
As a well known result [8], $\Sigma^{''} (\xi_{\bf k}, x)$ diverges
for $T \not= 0$.
Thus we may conclude that $\Omega_c$ diverges and this divergence
must be cancelled by a similar term in $\Omega_{d}$.
Now one may wonder whether there is any cancellation at the
self-energy level especially between the first and the second
diagrams in Fig.5 as in the case of the polarization bubbles.
Since the second diagram generates only the real part,
there is no cancellation in $\Sigma^{''}$.
For $\Sigma^{'}$, both of the two diagrams contribute.
However, one can see that there is no cancellation between the
two contributions because of the presence of the additional
fermion propagator in the first diagram.
We believe that these are the symptoms of the gauge non-invariant
nature of the self-energy.
In the previous sections, we consider first the polarization bubbles
which are gauge-invariant objects.
Note that there is an explicit cancellation in
this gauge-invariant combination.
Therefore, we think that using the polarization bubble makes the
gauge-invariance manifest.

Armed with these arguments, we can investigate the
$\Delta \omega_c \not= 0$ case.
Recalling that the first and the second terms of Eq.(12) correspond
to the first and the second diagrams of Fig.5, we may anticipate
a similar cancellation between these two terms as the case of
$\Delta \omega_c = 0$.
Indeed the f-sum rule, which is given by
$$
\sum_{lm} |M_{lm} ({\bf q} \rightarrow 0)|^2 \
{n_F (\xi_l) - n_F (\xi_m) \over \xi_{m} - \xi_{l}} =
{1 \over m} \left ( {m \Delta \omega_c \over 2 \pi}
\sum_{l} n_F (\xi_l) \right )  = {n_0 \over m} \ ,
\eqno{(37)}
$$
allows a cancellation between the first term and the
second term in the
$q \rightarrow 0$, $i \nu_n \rightarrow 0$ limit.
We use this result in the appendix to estimate various
contributions to the compressibility.

\vskip 0.5cm

\centerline{\bf V. CONCLUSION}

In the previous paper [17], we showed that the density-density
correlation function has a Fermi-liquid form as far as the long
wavelength and the low frequency limits are concerned.
An important issue is whether this result is compatible with the
previous self-consistency treatment based on the one-loop
self-energy correction [5] and the present calculation of the energy gap,
which are in favor of a divergent effective mass at the half-filling.
For a class of Fermi-liquid interaction parameters $f_{\bf pp'}$, which
gives a finite angular average $f_{0s}$, three of us with Furusaki
demonstrated that the
effective mass is finite if we want to fit the result of the
density-density correlation function to the usual
Fermi-liquid theory framework [17].
However, it is still possible that the effect of
the divergent effective mass is cancelled by a contribution
from a singular $f_{\bf pp'}$, which gives a divergent $f_{0s}$,
in the density-density correlation function [19].
{}From the previous paper [17], it is clear
that this scenario is possible only for $f_{\bf pp'} \propto
{1 \over |{\hat {\bf p}} - {\hat {\bf p}}'|}$ or $f_{\bf pp'} =
\zeta \delta ({\hat {\bf p}} - {\hat {\bf p}}')$ with a divergent
$\zeta$ (Note that an $f_{\bf pp'}$ of the delta function type
can be absorbed into the definition of the finite Fermi velocity
if $\zeta$ is finite [17]).
Even though this possibility is quite plausible, it is still not
clear whether we are allowed to interpret all physical
measurements in terms of the conventional Fermi-liquid theory.
Thus we are still at the stage of collecting necessary informations
for the ultimate understanding of the effect of the gauge field
fluctuations on the composite fermions.

Recently Stern and Halperin [19] calculated the energy gap
of the system from the one-particle Green's function of the
composite fermions in a finite effective magnetic field
$\Delta B$.
They identified the region where the spectral function vanishes at
zero temperature,
which is argued to be gauge-invariant, and found
an energy gap which is in agreement with the previous
self-consistency treatment [5] and the present calculation.
The advantage of our calculation is that we directly evaluated
the gauge-invariant two particle Green's function, and we
could consider the finite temperature situation.
We would like to mention that
the present perturbative calculation suggests
that the perturbation theory for the compressibility
breaks down for sufficiently small $\Delta \omega_c$
in the sense that the correction to the energy gap becomes
larger than the bare energy gap.
Thus one needs a truly gauge-invariant non-perturbative treatment
in order to understand this peculiar system.

\vskip 0.5cm

\centerline{\bf ACKNOWLEDGMENTS}

We would like to thank B. I. Halperin and Ady Stern for
valuable suggestions and important comments.
We are also grateful to Akira Furusaki for early
collaboration, and Manfred Sigrist for helpful discussions.
YBK and XGW are supported by NSF
grant No. DMR-9411574.
PAL is supported by NSF grant No. DMR-9216007.
PCES is supported by NSERC, and by an NSERC University Research
Fellowship.

\vfill\vfill\vfill
\break

\vskip 0.5cm

\centerline{\bigbf Appendix}

\vskip 0.5cm

In this appendix, we show that
${\partial n_{a2} \over \partial \mu}$ and
${\partial n_{b} \over \partial \mu}$ are exponentially smaller than
${\partial n_{a1} \over \partial \mu}$ which is calculated in the
main text.
As discussed in section IV, there is a partial cancellation between
$\Omega_{b1}$ and $\Omega_{b2}$ in Eq.(22) due to the f-sum rule
given by Eq.(37).
As a result, $\Omega_{b}$ can be rewritten as
$$
\eqalign{
\Omega_{b} &\approx \sum_{\bf q} \int^{\infty}_0 {dx \over \pi} \
(1 + 2 n_B (x)) \ {\widetilde D}^{''}_{11} ({\bf q}, x)
\left [ - \sum_{lm} |M_{lm} ({\bf q})|^2 \right. \cr
&\hskip 5.93cm \times \left.
\left ( {n_F (\xi_l) - n_F (\xi_m) \over x - \xi_m + \xi_l}
- {n_F (\xi_l) - n_F (\xi_m) \over  \xi_l - \xi_m} \right )
\right ] \  \cr
&= \sum_{\bf q} \int^{\infty}_0 {dx \over \pi} \
(1 + 2 n_B (x)) \ x \ {\widetilde D}^{''}_{11} ({\bf q}, x)
\left [ - \sum_{lm} |M_{lm} ({\bf q})|^2 \
{n_F (\xi_l) - n_F (\xi_m) \over (x - \xi_m + \xi_l) \ (\xi_l - \xi_m)}
\right ] \ .}
\eqno{\rm (A.1)}
$$

{}From Eq.(21) and Eq.(A.1), we get the lowest
order correction to the density of the fermions
$n_1 = n_a + n_b$ as follows.
$$
\eqalign{
n_a &= n_{a1} + n_{a2} \ , \cr
n_{a1} &= - {1 \over T} \sum_{\bf q} \sum_{l} |M_{ll} ({\bf q})|^2 \
D^{'}_{11} ({\bf q}, 0) \ n_F (\xi_l) (1 - n_F (\xi_l))
(1 - 2 n_F (\xi_l)) \ , \cr
n_{a2} &= - {1 \over T} \sum_{\bf q}
\sum_{l \not= m} |M_{lm} ({\bf q})|^2 \
D^{'}_{11} ({\bf q}, \xi_m - \xi_l) \
\left [ \ n_F (\xi_m) (1 - n_F (\xi_m)) (1 - n_F (\xi_l)) \right. \cr
&\hskip 6.2cm
- \left. n_F (\xi_m) n_F (\xi_l)) (1 - n_F (\xi_l)) \ \right ] \ ,}
\eqno{\rm (A.2)}
$$
and
$$
\eqalign{
n_{b} &\approx {1 \over T}
\sum_{\bf q} \int^{\infty}_0 {dx \over \pi} \
(1 + 2 n_B (x)) \ x \ D^{''}_{11} ({\bf q}, x) \cr
&\hskip 2.15cm \times
\left [ \sum_{lm} |M_{lm} ({\bf q})|^2 \ {n_F (\xi_l) (1 - n_F (\xi_l))
- n_F (\xi_m) (1 - n_F (\xi_m)) \over
(x - \xi_m + \xi_l) \ (\xi_l - \xi_m)} \right ] \ .}
\eqno{\rm (A.3)}
$$
These equations are equivalent to Eq.(17).

As shown in Eq.(18), in order to calculate the
compressibility, one should take the derivative of both
$D_{11} ({\bf q}, x)$ and $\Pi_{11} ({\bf q}, x)$.
Note that ${\partial D_{11} \over \partial \mu} \sim
D^{-2}_{11} {\partial \Pi_{11} \over \partial \mu}$ and
${\partial \Pi_{11} \over \partial \mu}$ contains the
factor $n_F (\xi_l) (1 - n_F (\xi_l))$.
Thus ${\partial D_{11} \over \partial \mu}$ generates additional
factors $e^{-|\xi_p|/T}$ and $e^{-\xi_{p+1}/T}$.
Since we want to keep only the terms which are proportional to
$e^{-|\xi_p|/T}$ or $e^{-\xi_{p+1}/T}$, we can ignore the terms
${\partial \Pi_{11} \over \partial \mu}
{\partial D_{11} \over \partial \mu}$, which are of order
$e^{-2|\xi_p|/T}$.
Ignoring these terms in Eq.(18) which is equivalent to keeping
only the $\mu$ dependence in $n_F$ in Eq.(A.2) and Eq.(A.3),
the lowest order correction
to the compressibility ${\partial n_1 \over \partial \mu}
= {\partial n_a \over \partial \mu} + {\partial n_b \over \partial \mu}$
can be calculated as follows.
$$
\eqalign{
{\partial n_a \over \partial \mu} &=
{\partial n_{a1} \over \partial \mu} +
{\partial n_{a2} \over \partial \mu} \ , \cr
{\partial n_{a1} \over \partial \mu} &\approx
- {1 \over T^2} \sum_{\bf q} \sum_{l} |M_{ll} ({\bf q})|^2 \
D^{'}_{11} ({\bf q}, 0) \cr
&\hskip 2.35cm \times n_F (\xi_l) (1 - n_F (\xi_l)) \left [ \
1 - 6 \ n_F (\xi_l) (1 - n_F (\xi_l)) \ \right ] \ , \cr
{\partial n_{a2} \over \partial \mu} &\approx
- {1 \over T^2} \sum_{\bf q} \sum_{l \not= m} |M_{lm} ({\bf q})|^2 \
D^{'}_{11} ({\bf q}, \xi_m - \xi_l) \cr
&\hskip 2.4cm \times \left [ \
n_F (\xi_m) (1 - n_F (\xi_m)) (1 - 2 n_F (\xi_m)) (1 - n_F (\xi_l))
\right. \cr
&\hskip 2.7cm - \left.
n_F (\xi_m) n_F (\xi_l) (1 - n_F (\xi_l)) (1 - 2 n_F (\xi_l))
\ \right ] \ ,}
\eqno{\rm (A.4)}
$$
and
$$
\eqalign{
{\partial n_{b} \over \partial \mu} &\approx
-{1 \over T^2}
\sum_{\bf q} \int^{\infty}_0 {dx \over \pi} \
(1 + 2 n_B (x)) \ x \ D^{''}_{11} ({\bf q}, x)
\Biggl [ \ \sum_{lm} {|M_{lm} ({\bf q})|^2 \over (x - \xi_m + \xi_l)
\ (\xi_l - \xi_m)} \cr
&\hskip 2.6cm \times
\biggl [ \ n_F (\xi_l) (1 - n_F (\xi_l)) (1 - 2 n_F (\xi_l)) \cr
&\hskip 3.3cm - n_F (\xi_m) (1 - n_F (\xi_m))
(1 - 2 n_F (\xi_m)) \ \biggr ]
\ \Biggr ] \ .}
\eqno{\rm (A.5)}
$$

Keeping only the terms that are proportional to $e^{-|\xi_p|/T}$
or $e^{-\xi_{p+1}/T}$, one can show that the contributions from
${\partial n_{a2} \over \partial \mu}$ and
${\partial n_{b} \over \partial \mu}$ do not contain such terms
that are proportional to $e^{-|\xi_p|/T}$ or $e^{-\xi_{p+1}/T}$.
This result can be obtained as follows.
In each case of ${\partial n_{a2} \over \partial \mu}$ and
${\partial n_{b} \over \partial \mu}$, the first term and the
second term inside the square bracket contain contributions
proportional to $e^{-|\xi_p|/T}$ or $e^{-\xi_{p+1}/T}$.
It can be seen that these contributions in the first term cancel
each other when the chemical potential lies exactly at the
middle of the successive effective Landau levels, and thus
they correspond to a uniform shift in these Landau levels.
The same story applies to the second term in the square bracket.
However, it turns out that the contributions from the first term
and the second term cancel again each other so that the
contributions proportional to $e^{-|\xi_p|/T}$ or $e^{-\xi_{p+1}/T}$
do not exist in general.
Thus ${\partial n_{a2} \over \partial \mu} =
{\cal O} (e^{-2|\xi_p|/T})$ and
${\partial n_{b} \over \partial \mu} =
{\cal O} (e^{-2|\xi_p|/T})$ so that we can ignore these
contributions compared to $e^{-|\xi_p|/T}$ or $e^{-\xi_{p+1}/T}$.

\vfill\vfill\vfill
\break

\vskip 0.5cm

\centerline{\bigbf References}

\vskip 0.5cm

\item{[1]} H. W. Jiang {\it et al.}, Phys. Rev. {\bf B} {\bf 40}, 12013 (1989).
\item{[2]} R. L. Willet {\it et al.}, Phys. Rev. Lett. {\bf 71}, 3846 (1993);
R. L. Willet {\it et al.}, Phys. Rev. {\bf B} {\bf 47}, 7344 (1993);
W. Kang {\it et al.}, Phys. Rev. Lett. {\bf 71}, 3850 (1993);
V. J. Goldman, B. Su, and J. K. Jain, Phys. Rev. Lett. {\bf 72},
2065 (1994).
\item{[3]} R. R. Du {\it et al.}, Phys. Rev. Lett. {\bf 70}, 2944 (1993);
D. R. Leadley {\it et al.}, Phys. Rev. Lett. {\bf 72}, 1906 (1994).
\item{[4]} J. K. Jain, Phys. Rev. Lett. {\bf 63}, 199 (1989);
Phys. Rev. {\bf B} {\bf 41}, 7653 (1990); Adv. Phys. {\bf 41}, 105 (1992).
\item{[5]} B. I. Halperin, P. A. Lee, and N. Read,
Phys. Rev. {\bf B} {\bf 47}, 7312 (1993).
\item{[6]} A. Lopez and E. Fradkin, Phys. Rev. {\bf B} {\bf 44},
5246 (1991); Phys. Rev. Lett. {\bf 69}, 2126 (1992).
\item{[7]} V. Kalmeyer and S. C. Zhang, Phys. Rev. {\bf B} {\bf 46},
9889 (1992).
\item{[8]} N. Nagaosa and P. A. Lee, Phys. Rev. Lett. {\bf 64}, 2550 (1990);
P. A. Lee and N. Nagaosa, Phys. Rev. {\bf B} {\bf 46}, 5621 (1992).
\item{[9]} B. L. Altshuler and L. B. Ioffe, Phys. Rev. Lett. {\bf 69},
2979 (1992).
\item{[10]} D. V. Khveshchenko, R. Hlubina, and T. M. Rice,
Phys. Rev. {\bf B} {\bf 48}, 10766 (1993).
\item{[11]} D. V. Khveshchenko and P. C. E. Stamp, Phys. Rev. Lett. {\bf 71},
2118 (1993); Phys. Rev. {\bf B} {\bf 49}, 5227 (1994).
\item{[12]} J. Polchinski, Nucl. Phys. {\bf B 422}, 617 (1994).
\item{[13]} Junwu Gan and Eugene Wong, Phys. Rev. Lett. {\bf 71},
4226 (1994).
\item{[14]} L. B. Ioffe, D. Lidsky, and B. L. Altshuler,
Phys. Rev. Lett. {\bf 73}, 472 (1994).
\item{[15]} H. -J. Kwon, A. Houghton, and J. B. Marston,
Phys. Rev. Lett. {\bf 73}, 284 (1994).
\item{[16]} C. Nayak, and F. Wilczek, Nucl. Phys. {\bf B 417}, 359 (1994).
\item{[17]} Y. B. Kim, A. Furusaki, X.-G. Wen, and P. A. Lee,
{\it Gauge-invariant response functions of fermions coupled to a gauge
field}, cond-mat/9405083.
\item{[18]} B. L. Altshuler, L. B. Ioffe, and A. J. Millis, {\it On the
low energy properties of fermions with singular
interactions}, cond-mat/9406024.
\item{[19]} A. Stern and B. I. Halperin, Private Communication.
\item{[20]} R. R. Du {\it et al.}, Solid State Comm. {\bf 90}, 71 (1994).
\item{[21]} S. He, P. M. Platzman, and B. I. Halperin,
Phys. Rev. Lett. {\bf 71}, 777 (1993);
Y. Hatsugai, P.-A. Bares, and X.-G. Wen,
Phys. Rev. Lett. {\bf 71}, 424 (1993);
Y. B. Kim and X.-G. Wen, Phys. Rev. {\bf B} {\bf 50},
8078 (1994).
\item{[22]} Y. H. Chen, F. Wilczek, E. Witten, and B. I. Halperin,
Int. J. Mod. Phys. {\bf 3}, 1001 (1989).
\item{[23]} S. H. Simon and B. I. Halperin, Phys. Rev. {\bf B} {\bf 48},
17368 (1993); S. H. Simon and B. I. Halperin, Phys. Rev. {\bf B} {\bf 50},
1807 (1994); Song He, S. H. Simon, and B. I. Halperin,
Phys. Rev. {\bf B} {\bf 50}, 1823 (1994).
\item{[24]} L. Zhang, {\it Finite temperature fractional
quantum Hall effect}, cond-mat/9409075.
\item{[25]} I. S. Gradshteyn and I. M. Ryzhik, {\it Table of Integrals,
Series, and Products}, Fifth Edition, Academic Press (1994).

\vfill\vfill\vfill
\break

\vskip 0.5cm

\centerline{\bigbf Figure captions}

\vskip 0.5cm

\item{Fig.1}
(a) The diagram that represents the density of the free fermions in an
effective magnetic field $\Delta B$.
(b) The lowest order correction to the density of the fermions due to the
gauge field fluctuation.
Here the solid line represents the bare electron propagator.
The wavy line denotes the RPA gauge field propagator which is given by
the diagram in Fig.2 (a).

\vskip 0.5cm

\item{Fig.2}
(a) The wavy line denotes the RPA gauge field propagator
and the dashed line is the bare gauge field propagator.
Here the hatched bubble (b) represents the transverse part
of the polarization bubble.

\vskip 0.5cm

\item{Fig.3}
The diagrams that correspond to the thermodynamic potential of the
free fermions (a) and the gauge field contribution (b) to the
thermodynamic potential.

\vskip 0.5cm

\item{Fig.4}
The diagrams that represent the lowest order correction to the
compressibility of the fermions.

\vskip 0.5cm

\item{Fig.5}
The diagrams that represent the lowest order correction to the
self-energy of the fermions.

\bye